\documentclass[runningheads]{llncs}
\usepackage{amsmath}
\usepackage{amssymb}
\usepackage{graphicx}
\usepackage{hyperref}
\usepackage{breakurl}
\usepackage{url}
\usepackage{color}
\usepackage[strings]{underscore}

\begin{document}
\title{FunGrim: a symbolic library for special functions}
%
%
\author{Fredrik Johansson \orcidID{0000-0002-7368-092X}}
\authorrunning{F. Johansson}

\institute{LFANT, Inria Bordeaux, Talence, France \\
\email{fredrik.johansson@gmail.com}\\
\url{http://fredrikj.net}}
\maketitle              
\begin{abstract}
We present the Mathematical Functions Grimoire (FunGrim), a
website and database of formulas and theorems for special functions.
We also discuss the symbolic computation library
used as the backend and main development tool for FunGrim,
and the Grim formula language used in these projects
to represent mathematical content semantically.

\end{abstract}
\section{Introduction}

The \emph{Mathematical Functions Grimoire}\footnote{A \emph{grimoire} is a book of magic formulas.}
(FunGrim, \url{http://fungrim.org/}) is
an open source library of formulas, theorems and data for mathematical functions.
It currently contains around 2600 entries. As one example entry,
the modular transformation law
of the Eisenstein series $G_{2k}$ on the upper half-plane $\mathbb{H}$
is given in \url{http://fungrim.org/entry/0b5b04/} as follows:
$$G_{2 k}\!\left(\frac{a \tau + b}{c \tau + d}\right) = {\left(c \tau + d\right)}^{2 k} G_{2 k}\!\left(\tau\right)$$
\begin{center}
Assumptions: $k \in \mathbb{Z}_{\ge 2} \;\mathbin{\operatorname{and}}\; \tau \in \mathbb{H} \;\mathbin{\operatorname{and}}\; \begin{pmatrix} a & b \\ c & d \end{pmatrix} \in \operatorname{SL}_2(\mathbb{Z})$
\end{center}

FunGrim stores entries as symbolic expressions with metadata, in this case:

\begin{small}
\begin{verbatim}
Entry(ID("0b5b04"),
    Formula(Equal(EisensteinG(2*k, (a*tau+b)/(c*tau+d)),
        (c*tau+d)**(2*k) * EisensteinG(2*k, tau))),
    Variables(k, tau, a, b, c, d),
    Assumptions(And(Element(k, ZZGreaterEqual(2)), Element(tau, HH),
        Element(Matrix2x2(a, b, c, d), SL2Z))))
\end{verbatim}
\end{small}

Formulas are fully quantified (\emph{assumptions} give conditions for the free variables
such that the formula is valid)
and context-free (symbols have a globally consistent meaning), giving
precise statements of mathematical theorems.
The metadata may also include bibliographical references.
Being easily computer-readable, the database may be used
for automatic term rewriting in symbolic algorithms.
This short paper discusses the semantic representation
of mathematics in FunGrim and the underlying software.

\section{Related projects}

FunGrim is in part a software project and in part
a reference work for mathematical functions in the tradition of Abramowitz and Stegun~\cite{AbramowitzStegun1964}
but with updated content and a modern interface.
There are many such efforts, notably
the NIST Digital Library of Mathematical Functions (DLMF)~\cite{NIST:DLMF}
and the Wolfram Functions Site (WFS)~\cite{WolframFunctions}, which have two rather different approaches:

\begin{itemize}
\item DLMF uses LaTeX together with prose for its content.
Since many formulas depend on implicit context and LaTeX is presentation-oriented
rather than semantic (although DLMF adds semantic extensions to LaTeX
to alleviate this problem),
the content is not fully computer-readable and
can also sometimes be ambiguous to human readers.
DLMF is edited for conciseness, giving
an overview of the main concepts
and omitting in-depth content.
\item WFS represents the content as context-free
symbolic expressions written in the Wolfram Language.
The formulas can be parsed by Mathematica,
whose evaluation semantics provide concrete meaning.
Most formulas are computer-generated, sometimes exhaustively
(for example, WFS lists tens of thousands of
transformations between elementary functions and around 200,000
formulas for special cases of hypergeometric functions).
\end{itemize}

FunGrim uses a similar approach to that of WFS, but does not depend on
the proprietary Wolfram technology.
Indeed, one of the central reasons for starting FunGrim
is that both DLMF and WFS are not open source (though freely accessible).
Another central idea behind FunGrim is to provide even stronger semantic
guarantees; this aspect is discussed in a later section.

Part of the motivation is also to offer complementary content:
in the author's experience, the DLMF and WFS are strong in some
areas and weak in others. For example,
both have minimal coverage of some important functions
of number theory and
they cover \emph{inequalities} far less extensively than \emph{equalities}.
At this time, FunGrim has perhaps 10\% of the content needed for
a good general reference on special functions, but as proof as concept,
it has detailed content for some previously-neglected topics.
The reader may compare the following:
\begin{itemize}
\item {\small \url{http://fungrim.org/topic/Modular_lambda_function/}} versus \\
{\small \url{http://functions.wolfram.com/EllipticFunctions/ModularLambda/}} versus \\
formulas for $\lambda(\tau)$ in {\small \url{https://dlmf.nist.gov/23.15}} + {\small \url{https://dlmf.nist.gov/23.17}}. \vskip2pt
\item {\small \url{http://fungrim.org/topic/Barnes_G-function/}} versus \\ {\small \url{https://dlmf.nist.gov/5.17}}. (The Barnes G-function is not covered in WFS.)
\end{itemize}

Most FunGrim content is hand-written so far; adding
computer-generated
entries in the same fashion as WFS is a future possibility.

We mention three other related projects:

\begin{itemize}

\item FunGrim shares many goals with 
the NIST Digital Repository of Mathematical Formulas (DRMF)~\cite{cohl2014digital},
a companion project to the DLMF.
We will not attempt to compare the projects in depth since
DRMF is not fully developed, but we mention one important
difference: DRMF represents formulas using
a semantic form of LaTeX which is hard
to translate perfectly to symbolic expressions,
whereas FunGrim (like WFS) uses symbolic expressions as the
source representation and generates LaTeX automatically
for presentation.

\item The Dynamic Dictionary of Mathematical
Functions (DDMF)~\cite{benoit2010dynamic} generates
information about mathematical functions algorithmically,
starting \emph{ab initio} only from the defining differential equation of each function.
This has many advantages: it enables a high degree of reliability
(human error is removed from the equation, so to speak),
the presentation is uniform, and it is easy to add new functions.
The downside is that the approach is limited to a restricted class
of properties for a restricted class of functions.

\item The LMFDB~\cite{lmfdb} is a large database of
L-functions, modular forms, and related objects.
The content largely consists of data tables
and does not include ``free-form'' symbolic formulas and theorems.

\end{itemize}

\section{Grim formula language}

\emph{Grim} is the symbolic mathematical language used in FunGrim.\footnote{Documentation of the Grim language is available at \url{http://fungrim.org/grim/}}
Grim is designed to be easy to write and parse
and to be embeddable within a host programming language such as Python,
Julia or JavaScript
using the host language's native syntax (similar to SymPy~\cite{10.7717/peerj-cs.103}).
The reference implementation is Pygrim,
a Python library which implements Grim-to-LaTeX conversion and
symbolic evaluation of Grim expressions.
Formulas are converted to HTML using KaTeX for display on the FunGrim website;
Pygrim also provides hooks to show Grim expressions as LaTeX-rendered formulas
in Jupyter notebooks.
The FunGrim database itself is currently part of the Pygrim source code.\footnote{Pygrim is currently in early development
and does not have an official release. The source
code is publicly available at \url{https://github.com/fredrik-johansson/fungrim}}

Grim has a minimal core language,
similar to Lisp S-expressions and Wolfram language M-expressions.
The only data structure is an expression tree composed of function
calls \texttt{f(x, y, ...)} and atoms
(integer literals, string literals,
alphanumerical symbol names). For example, \texttt{Mul(2, Add(a, b))} represents $2 (a+b)$.
For convenience, Pygrim uses operator overloading in Python
so that the same expression may be written more simply as \texttt{2*(a+b)}.

On top of the core language, Grim provides a
vocabulary of hundreds of builtin symbols (\texttt{For}, \texttt{Exists}, \texttt{Matrix}, \texttt{Sin}, \texttt{Integral}, etc.)
for variable-binding, logical operations, structures,
mathematical functions, calculus operations, etc.

The following dummy formula is a more elaborate example:

\begin{small}
\begin{verbatim}
Where(Sum(1/f(n), For(n, -N, N), NotEqual(n, 0)), Def(f(n),
    Cases(Tuple(n**2, CongruentMod(n, 0, 3)), Tuple(1, Otherwise))))
\end{verbatim}
\end{small}

$$\sum_{\textstyle{n=-N \atop n \ne 0}}^{N} \frac{1}{f(n)}\; \text{ where } f(n) = \begin{cases} {n}^{2}, & n \equiv 0 \pmod {3}\\1, & \text{otherwise}\\ \end{cases}$$

Grim can be used both as a mathematical markup language and as a simple functional
programming language. 
Its design is deliberately constrained:

\begin{itemize}
\item Grim is not intended to be a typesetting language: the Grim-to-LaTeX converter takes care of
most presentation details automatically. (The results are not always perfect, and Grim
does allow including typesetting hints where the default rendering is inadequate.)

\item Grim is not intended to be a general-purpose programming language.
Unlike full-blown Lisp-like programming languages, Grim is not meant to be used to manipulate symbolic expressions from within,
and it lacks concrete data structures for programming, being mainly concerned with representing immutable mathematical objects. 
Grim is rather meant to be embedded in a host programming language where the host language can be used to traverse expression trees
or implement complex algorithms.

\end{itemize}

Grim formulas entered in Pygrim are preserved verbatim
until explicitly evaluated.
This contrasts with most computer algebra systems, which automatically
convert expressions to ``canonical'' form.
For example, SymPy automatically rewrites $2(b+a)$ as $2a+2b$ (distributing 
he numerical coefficient
and sorting the terms).
SymPy's behavior can be overridden with a special ``hold'' command,
but this can be a hassle to use and might not be
recognized by all functions.

\section{Evaluation semantics}

FunGrim and the Grim language have the following fundamental semantic rules:
\begin{itemize}
\item Every mathematical object or operator must have an unambiguous
interpretation, which cannot vary with context.
In principle, every syntactically valid
constant expression should represent a definitive
mathematical object (possibly the special object $\text{Undefined}$
when a function is evaluated outside its domain of definition).
This means, for example, that multivalued functions have fixed
branch cuts (analytic continuation must be expressed explicitly), and
removable singularities do not cancel automatically.
Many symbols which have an overloaded meaning in standard mathematical
notation require disambiguation; for example, Grim provides separate \texttt{SequenceLimit},
\texttt{RealLimit} and \texttt{ComplexLimit} operators to express $\lim_{x\to c} f(x)$,
depending on whether the set of approach is meant as $\mathbb{Z}$, $\mathbb{R}$ or $\mathbb{C}$.
\vskip5pt

\item The standard logical and set operators ($=$ and $\in$, etc.) compare identity of mathematical objects, not equivalence under morphisms.
The mathematical universe is constructed to have few, orthogonal ``types'': for example,
the integer 1 and the complex number 1 are the same object, with $\mathbb{Z} \subset \mathbb{C}$. \vskip5pt

\item Symbolic evaluation (rewriting an expression as a simpler expression, e.g.\ $2 + 2 \rightarrow 4$)
must preserve the exact value of the input expression.
Formulas containing free variables are implicitly quantified
over the whole universe unless explicit assumptions are provided,
and may only be rewritten in ways that preserve the value
for all admissible values of the free variables.
For example, $yx \rightarrow xy$ is not a valid rewrite operation \emph{a priori}
since the universe contains noncommutative objects such as matrices,
but it is valid when quantified with assumptions that make $x$ and $y$ commute,
e.g.\ $x, y \in \mathbb{C}$.
\end{itemize}

These semantics are stronger than in most symbolic computing environments.
Computer algebra systems traditionally ignore
``exceptional cases'' when rewriting expressions.
For example, many computer algebra systems automatically simplify $x / x$ to $1$,
ignoring the exceptional case $x = 0$ where a division by zero occurs.\footnote{The simplification is valid if $x$ is viewed as a formal indeterminate
generating $\mathbb{C}[x]$ rather than a free variable representing a complex number. The point remains that some computer algebra
systems overload variables to serve both purposes, and this ambiguity is a frequent source of bugs. In Grim, the distinction is explicit.}
A more extreme example is to blindly simplify $\sqrt{x^2} \rightarrow x$
(invalid for negative numbers), and more generally to ignore
branch cuts or complex values.

Indeed, one section of the Wolfram Mathematica documentation helpfully warns users:
``The answer might not be valid for certain exceptional values of the parameters.''
As a concrete illustration, we can use Mathematica to
``prove'' that $e = 2$ by evaluating the hypergeometric function ${}_1F_1(a,b,1)$ at $a=b=-1$ using
two different sequences of substitutions:

\begin{itemize}
\item ${}_1F_1(a, b, 1) \; \rightarrow \; [a = b] \; \rightarrow \;  e \; \rightarrow \;  [b = -1] \; \rightarrow \;  e$
\item ${}_1F_1(a, b, 1) \; \rightarrow \;  [a = -1] \; \rightarrow \;  1-\frac{1}{b} \; \rightarrow \;  [b = -1] \; \rightarrow \;  2$
\end{itemize}

The contradiction happens because Mathematica uses two different
rules to rewrite the ${}_1F_1$ function, and the rules are inconsistent
with each other in the exceptional case $a = b \in \mathbb{Z}_{\le 0}$).\footnote{In WFS, corresponding
contradictory formulas are
\url{http://functions.wolfram.com/07.20.03.0002.01} and \url{http://functions.wolfram.com/07.20.03.0118.01}.}
(SymPy has the same issue.)

Our aspiration for the Grim formula language
and the FunGrim database is to make such contradictions impossible
through strong semantics and pedantic use of assumptions.
This should aid human understanding
(a user can inspect the source code of a formula and look
up the definitions of the symbols) and help support
symbolic computation, automated testing,
and possibly formal theorem-proving efforts.
Perfect consistency is particularly important for working with multivariate functions,
where corner cases can be extremely difficult to spot.

In reality, eliminating inconsistencies is an asymptotic goal:
there are certainly present and future mathematical errors in the FunGrim database
and bugs in the Pygrim reference implementation. We believe that
such errors can be minimized through randomized testing
(ideally combined with formal verification in the future, where such methods are applicable).

\section{Evaluation with Pygrim}

Pygrim has rudimentary support for evaluating and simplifying Grim expressions.
It is able to perform basic logical and arithmetic operations,
expand special cases of mathematical functions,
perform simple domain inferences,
partially simplify symbolic arithmetic expressions,
evaluate and compare algebraic numbers using an exact
implementation of $\overline{\mathbb{Q}}$ arithmetic,
and compare real or complex numbers using Arb enclosures~\cite{Johansson2017arb}
(only comparisons of unequal numbers can be decided in this way; equal numbers
have overlapping enclosures and can only be compared
conclusively when an algebraic or symbolic simplification is possible).

Calling the \texttt{.eval()} method in Pygrim returns an evaluated expression:

\begin{small}
\begin{verbatim}
>>> Element(Pi, SetMinus(OpenInterval(3, 4), QQ)).eval()
True_

>>> Zeros(x**5 - x**4 - 4*x**3 + 4*x**2 + 2*x - 2,
...           ForElement(x, CC), Greater(Re(x), 0)).eval()
...
Set(Sqrt(Add(2, Sqrt(2))), 1, Sqrt(Sub(2, Sqrt(2))))

>>> ((DedekindEta(1 + Sqrt(-1)) / Gamma(Div(5, 4))) ** 12).eval()
Div(-4096, Pow(Pi, 9))
\end{verbatim}
\end{small}

To simplify formulas involving free variables, the user needs
to supply sufficient assumptions:

\begin{small}
\begin{verbatim}
>>> (x / x).eval()
Div(x, x)
>>> (x / x).eval(assumptions=Element(x, CC))
Div(x, x)
>>> (x / x).eval(assumptions=And(Element(x, CC), NotEqual(x, 0)))
1
>>> Sin(Pi * n).eval()
Sin(Mul(Pi, n))
>>> Sin(Pi * n).eval(assumptions=Element(n, ZZ))
0
\end{verbatim}
\end{small}

In some cases, Pygrim can output conditional expressions: for example,
the evaluation
${}_2F_1(1, 1, 2, x) = -\log(1-x)/x$ is made with an explicit case distinction
for the removable singularity at $x = 0$ (the singularity
at $x = 1$ is consistent with $\log(0) = -\infty$ and does not
require a case distinction).

\begin{small}
\begin{verbatim}
>>> f = Hypergeometric2F1(1, 1, 2, x); f.eval()
Hypergeometric2F1(1, 1, 2, x)            # no domain -- no evaluation
>>> f.eval(assumptions=Element(x, CC))
Cases(Tuple(Div(Neg(Log(Sub(1, x))), x), NotEqual(x, 0)),
    Tuple(1, Equal(x, 0)))               # separate case for x = 0
>>> f.eval(assumptions=Element(x, SetMinus(CC, Set(0))))
Div(Neg(Log(Sub(1, x))), x)              # no case distinction needed
\end{verbatim}
\end{small}

Pygrim is not a complete computer algebra system; its features
are tailored to developing FunGrim and exploring
special function identities. Users
may also find it interesting as a symbolic interface to Arb (the
\texttt{.n()} method returns an arbitrary-precision enclosure of a constant expression).

\section{Testing formulas}

To test a formula $P(x_1,\ldots,x_n)$ with free variables
$x_1,\ldots,x_n$ and corresponding assumptions $Q(x_1,\ldots,x_n)$, we
generate pseudorandom values $x_1,\ldots,x_n$ satisfying $Q(x_1,\ldots,x_n)$,
and for each such assignment we evaluate the constant expression $P(x_1,\ldots,x_n)$.
If $P$ evaluates to False, the test fails (a counterexample has been found).
If $P$ evaluates to True or cannot be simplified to True/False (the truth value is unknown),
the test instance passes.

As an example, we test $P(x) = [\sqrt{x^2} = x]$
with assumptions $Q(x) = [x \in \mathbb{R}]$:
\begin{small}
\begin{verbatim}
>>> formula = Equal(Sqrt(x**2), x)
>>> formula.test(variables=[x], assumptions=Element(x, RR))
{x: 0}    ...  True
{x: Div(1, 2)}    ...  True
{x: Sqrt(2)}    ...  True
{x: Pi}    ...  True
{x: 1}    ...  True
{x: Neg(Div(1, 2))}    ...  False
\end{verbatim}
\end{small}

The test passes for $x = 0, \tfrac{1}{2}, \sqrt{2}, \pi, 1$, but $x = -\tfrac{1}{2}$
is a counterexample. With correct assumptions $x \in \mathbb{C} \,\land\, \left(\operatorname{Re}(x) > 0 \,\lor\, \left(\operatorname{Re}(x) = 0 \,\land\, \operatorname{Im}(x) > 0\right)\right)$, it passes:

\begin{small}
\begin{verbatim}
>>> formula.test(variables=[x], assumptions=And(Element(x, CC),
...     Or(Greater(Re(x), 0), And(Equal(Re(x), 0), Greater(Im(x), 0)))))
...
Passed 77 instances (77 True, 14 Unknown, 0 False)
\end{verbatim}
\end{small}

It currently takes two CPU hours to test the FunGrim database
with up to 100 test instances (assignments $x_1,\ldots,x_n$
that satisfy the assumptions) per entry.
We estimate that around 75\% of the entries are effectively testable.
For the other 25\%, either the symbolic evaluation code in Pygrim
is not powerful enough to generate
any admissible values (for which $Q$ is provably True),
or $P$ contains constructs
for which Pygrim does not yet support symbolic or numerical evaluation.
For 30\% of the entries, Pygrim is able to symbolically simplify
$P$ to True in at least one test instance (in the majority of cases,
it is only able to check consistency via Arb).
We aim to improve all these statistics in the future.

The test strategy is effective: the first run to test the FunGrim database found errors in 24 out of 2618
entries. Of these, 4 were mathematically wrong formulas
(for example, the Bernoulli number inequality ${\left(-1\right)}^{n} B_{2 n + 2} > 0$
had the prefactor negated as $(-1)^{n+1}$), 6 had incorrect
assumptions (for example, the Lambert W-function identity $W_{0}\!\left(x \log(x)\right) = \log(x)$ was given with assumptions $x \in [-e^{-1},\infty)$
instead of the correct $x \in [e^{-1},\infty)$); the remaining errors were
due to incorrect metadata or improperly constructed symbolic expressions.

A similar number of additional errors were found
and corrected after improving Pygrim's evaluation code further.
An error rate near 5\% seems plausible for untested formulas entered by hand
(by this author!).
We did not specifically search for errors in the literature used as
reference material for FunGrim; however, many corrections
were naturally made when the entries were first added,
prior to the development of the test framework.

\section{Formulas as rewrite rules}

The FunGrim database can be used for term rewriting, most easily by
applying a specific entry as a rewrite rule.
For example, FunGrim entry \texttt{ad6c1c} is the trigonometric identity
$\sin(a) \sin(b) = \tfrac{1}{2}\left(\cos\left(a - b\right) - \cos\left(a + b\right)\right)$:

\begin{small}
\begin{verbatim}
>>> (Sin(2) * Sin(Sqrt(2))).rewrite_fungrim("ad6c1c")
Div(Sub(Cos(Sub(2, Sqrt(2))), Cos(Add(2, Sqrt(2)))), 2)
\end{verbatim}
\end{small}

This depends on pattern matching.
To ensure correctness, a match is only made if
parameters in the input expression satisfy
the assumptions for free variables listed in the FunGrim entry.
The pattern matching is currently implemented naively
and will fail to match expressions that are mathematically equivalent
but structurally different
(better implementations are possible~\cite{krebber2018matchpy}).

A rather interesting idea is to search the whole database automatically
for rules to apply to simplify a given formula. We have used this
successfully on toy examples, but much more work is needed to develop
a useful general-purpose simplification engine; this would require stronger
pattern matching as well as heuristics for applying sequences
of rewrite rules. Rewriting using a database is perhaps most likely to be successful
for specific tasks and in combination with advanced hand-written search heuristics
(or heuristics generated via machine learning).
A prominent example of the hand-written approach
is Rubi~\cite{rich2018rule} which uses a
decision tree of thousands of rewrite rules
to simplify indefinite integrals.

\bibliographystyle{splncs04}
\bibliography{mybibliography}

\end{document}